\documentclass{PoS}
\usepackage{epsfig}
\usepackage{amsmath} 
\usepackage{amssymb} 
\usepackage{lineno}
\usepackage{graphicx}

\title{Volume Effects in Discrete $\beta$ functions}

\ShortTitle{Volume Effects in Discrete $\beta$ functions}

\author{\speaker{Yuzhi Liu}, Y. Meurice, and Haiyuan Zou\\
        Department of Physics and Astronomy\\The University of Iowa\\
        E-mail: \email{yuzhi-liu@uiowa.edu}\\
}

\abstract{We calculate discrete beta functions corresponding to the two-lattice matching for the $2D$ $O(N)$ models and Dyson's hierarchical model. We describe and explain finite-size effects such as the appearance of a nontrivial infrared fixed point that goes to infinity at infinite volume or the merging of an infrared and an ultraviolet fixed point.  We present extensions of the RG flows to the complex coupling plane. We discuss the possibility of constructing a continuous beta function from the discrete one by using functional conjugation methods. We briefly discuss the relevance of these findings for the search of nontrivial fixed points in multiflavor lattice gauge theory models.}

\FullConference{ The XXIX International Symposium on Lattice Field Theory - Lattice 2011\\
July 10-16, 2011\\
Squaw Valley, Lake Tahoe, California}

\begin{document}

\section{Introduction and Background}

With the upcoming experimental effort at the Large Hadron Collider (LHC), there has been a lot
of activity in the lattice community regarding the existence of non-trivial infrared
fixed points (IRFP) \cite{DeGrand:2010ba}. Different methods have been used to look for or exclude IRFP and some results are controversial. Calculating the discrete $\beta$ function is one of the methods used. In the following, we discuss finite volume effects in discrete $\beta$ functions for spin models that can be solved by numerical methods. 

\section{ $\beta$ functions and Two Lattice Matching}

In the continuum, the  Callan-Symanzik $\beta$ function, $\beta_{CS}$, can be calculated using standard perturbation theory
 
\begin{equation}
\beta_{CS}(\alpha)=-\beta_1 \alpha^2-\beta_2 \alpha^3-\beta_3 \alpha^4+\cdots
\label{eq:beta_cs}
\end{equation}
where $\alpha=\frac{g^2}{4\pi}$. The first two coefficients of Eq. \ref{eq:beta_cs} are regularization scheme independent. They are functions of the number of colors $N_c$ and the number of quark flavors $N_f$ \cite{PhysRevLett.33.244,1974NuPhB..75..531J}:
\begin{align}
\beta_1 &= \frac{1}{6\pi}(11N_c-2N_f)\\
\beta_2 &= \frac{1}{24\pi^2}(34N_c^2-10N_cN_f-3\frac{N_c^2-1}{N_c}N_f)
\end{align}
The theory is asymptotically free if $\beta_1>0$ and a stable IRFP may emerge for $\beta_1>0$ and $\beta_2<0$. The IRFP may appear in the large $\alpha$ region where perturbation theory breaks down. Therefore, it is essential to locate IRFP nonperturbatively. 

One method to calculate a discrete $\beta$ function nonperturbatively is called the two lattice matching \cite{1984PhLB..140...76H,1988PhLB..211..132G}. It is a way of measuring the running of bare couplings based on the fact that all the observables will have the same value if the models have the same effective action. One can block-spin the system of volume $V$, where $V$ is the volume in lattice units, $n$ times and calculate an observable that will be denoted $R(g,V)$. One can also block-spin the system of smaller volume $V/b^D$ $n-1$ times, calculate the observable $R(g^\prime, V/b^D)$, and tune the new coupling $g^\prime$ until the two observables match. If all the observables give the same value for these two coupling configurations $g$ and $g^\prime$, then one the two blocked system have the same large distance physics and the same physical correlation length. In the models discussed below, we only consider one parameter in the coupling space. In general, many more couplings are generated during the block-spin transformation. If only one relevant parameter exists, all the other irrelevant couplings will eventually die out as one block-spins many times near the critical point. Therefore, the examples showed below should apply to a larger class of system. In Ref.  \cite{PhysRevD.83.096008}, we used the following observable: 
\begin{equation}
R(\beta,\mathcal{V}/a^D)\equiv\frac{\left\langle (\sum_{x\in B_1}\vec{\phi}_x )(\sum_{y\in B_2}\vec{\phi}_y)\right\rangle_\beta}{\left\langle (\sum_{x\in B_1}\vec{\phi}_x)(\sum_{y\in B_1}\vec{\phi}_y ))\right\rangle_\beta} 
\end{equation}
where $\mathcal{V}$ is the physical volume of the system, $a$ is the lattice spacing, $B_1$ and $B_2$ denote neighbor blocks and $D$ is the dimensionality of the system. The reason we chose this particular rational form is that we do not need to calculate the partition function explicitly since it cancels out in both denominator and numerator. 
The matching condition reads $R(\beta,L^D)=R(\beta',(L/b)^D)$, where $b$ is the scaling factor. The discrete $\beta$ function is defined as $\Delta\beta(\beta,L^D\rightarrow (L/b)^D)= \beta - \beta^\prime$. This definition is consistent with the one used in \cite{1984PhLB..140...76H,1988PhLB..211..132G}. $\beta$ denotes either a quantity proportional to $g^{-2}$ or the inverse temperature and should not be confused with $\beta_{CS}$. In order to calculate $\Delta\beta(\beta_0)$, we start out from large volume $L^D$ at $\beta_0$ and tune the coupling at smaller volume $(L/b)^D$ to $\beta_1$ so that $R(\beta_0,L^D)=R(\beta_1,(L/b)^D)$. We then match $R(\beta_1,L^D)=R(\beta_2,(L/b)^D)$ and so on. By repeating this procedure, we get a sequence of $\beta$'s from which we can calculate the discrete $\beta$ function defined above. In the following, we will work with Dyson's hierarchical Ising model and the $D=2$ nonlinear $O(N)$ sigma model in the large $N$ limit.

\section{Dyson's Hierarchical Ising Model}

The Hamiltonian of the Dyson's hierarchical model with $2^{n_{max}}$ lattice sites is defined as 
\begin{equation}
H =
-{\frac{1}{2}}\sum_{n=1}^{n_{max}}({\frac{c}{4}})^n f(n) \sum_{B^{(n)}}
(\sum  _{x \in  B^{(n)} } \phi_x)^{2} \ 
\label{eq:Hamiltonian}
\end{equation}
where $c$ controls the interaction strength for different block sizes. We can include the dimensionality $D$ through the relation $c/4=b^{-2-D}$ \cite{2007JPhA...40...39M}, where $b=2^{1/D}$ is the scaling factor. 
The model has many nice properties and is an ideal laboratory to test various ideas before one applies them to the more complicated full QCD case. Here is a list of properties we used \cite{1969CMaPh..12...91D,1971CMaPh..21..269D,pool1972mathematical}
\begin{itemize}
\item For $D>2$ and $f(n)=1$, the model has a second order phase transition, which is similar to the $D=3$ regular (nearest neighbor) Ising model. 
\item For $D\leq 2$ and $f(n)=1$, the model has no phase transition at finite temperature, which is different from the  $D=2$ regular Ising model.
\item For $D=2$ and $f(n)=log(n)$, the model is equivalent to Anderson model and has a Thouless effect (discontinuity in the magnetization). 
\end{itemize}
The cases $D$= 3, 2, 1.7 with $f(n)=1$  have been analyzed thoroughly in \cite{PhysRevD.83.096008}. In the following, we will consider various dimensions for $f(n)=log(n+1)$ ( models with $f(n)=log(n+1)$ have same properties as $f(n)=log(n)$ in the infinite volume limit).

\section{How does an IRFP appear/disappear?}

\begin{figure}
\begin{center}
\hspace{-0.03\textwidth}
\includegraphics[width=0.35\textwidth,keepaspectratio=]{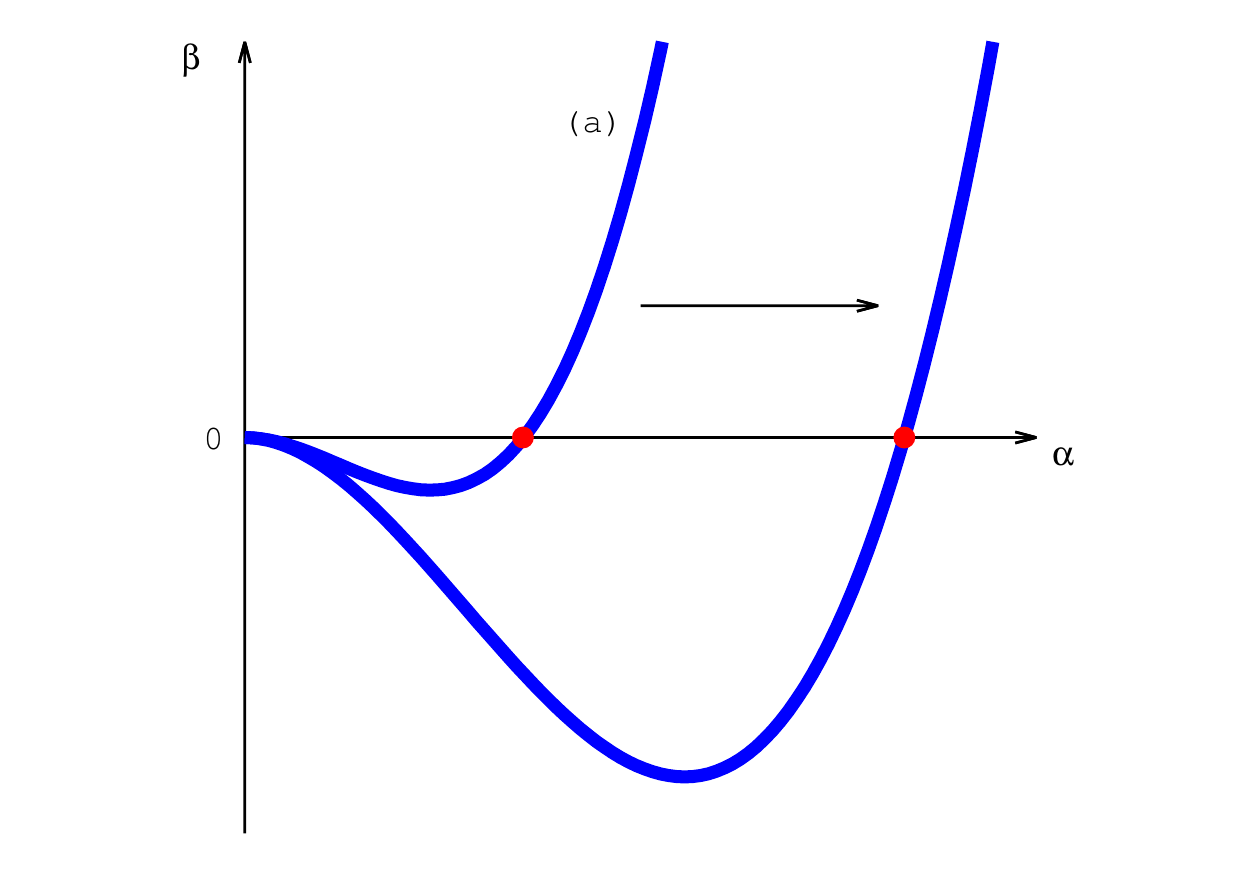}
\hspace{-0.03\textwidth}
\includegraphics[width=0.35\textwidth,keepaspectratio=]{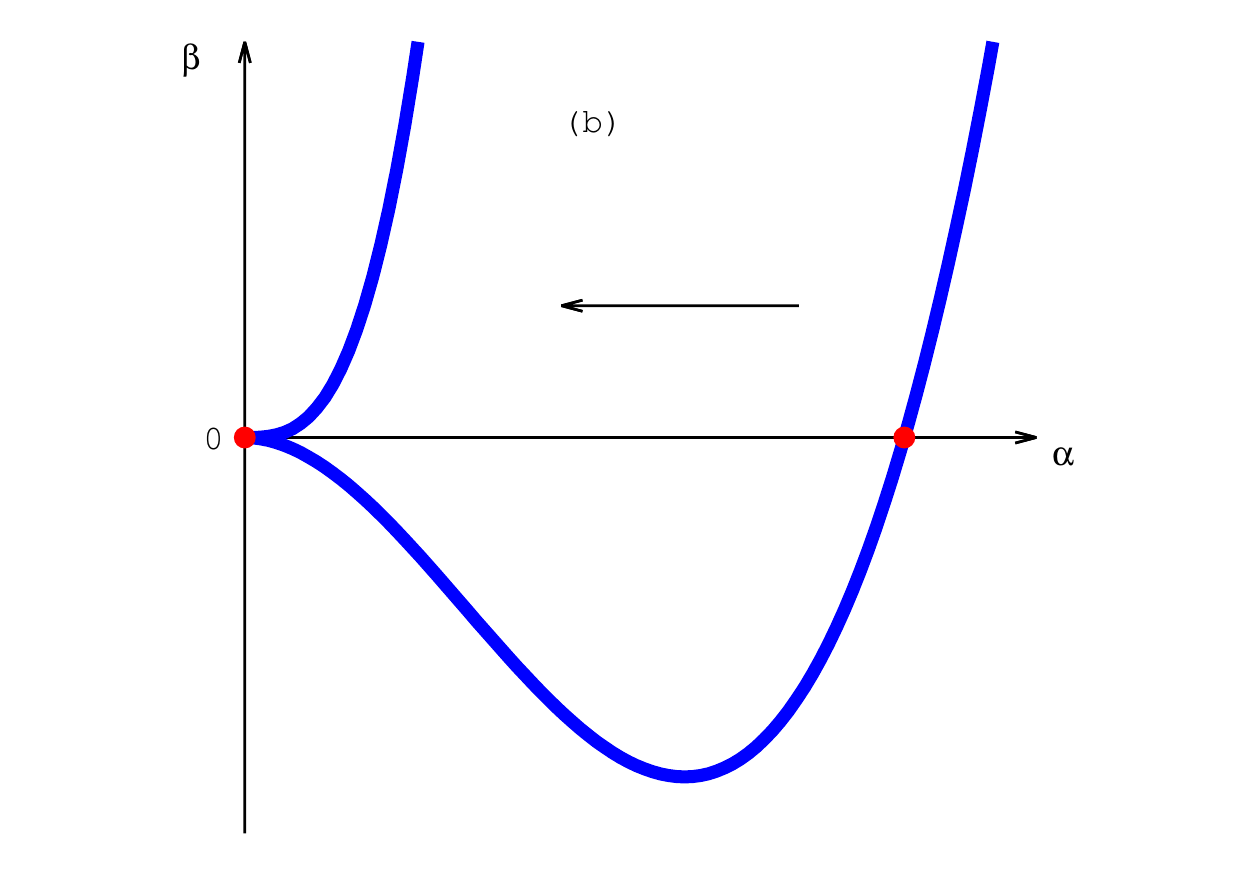}
\hspace{-0.03\textwidth}
\includegraphics[width=0.35\textwidth,keepaspectratio=]{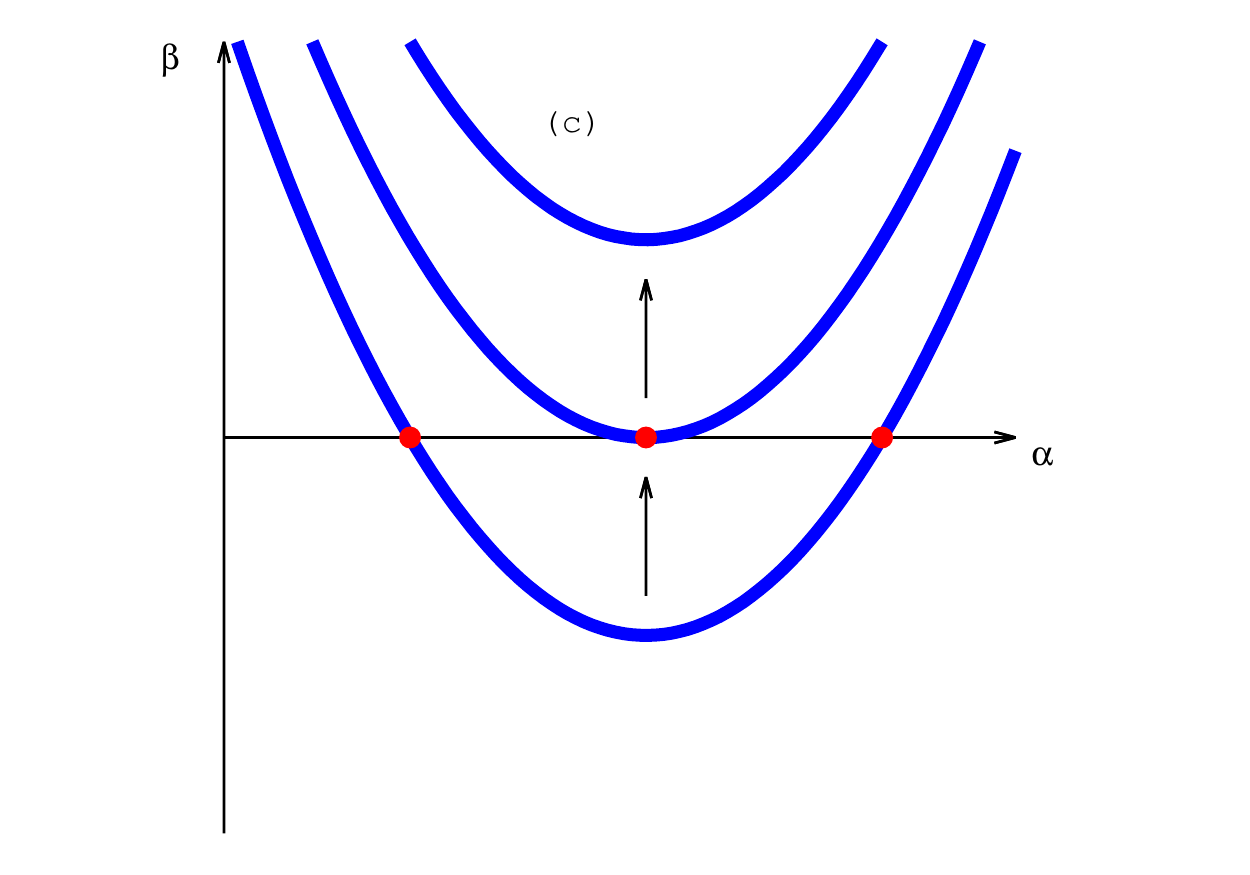}
\end{center}
\caption{Three mechanisms for the loss of the fixed point(s) described in \cite{PhysRevD.80.125005}. }
\label{fig:zeros_kaplan}
\end{figure}

\begin{figure}[htp]
\begin{center}
\includegraphics[height=7cm]{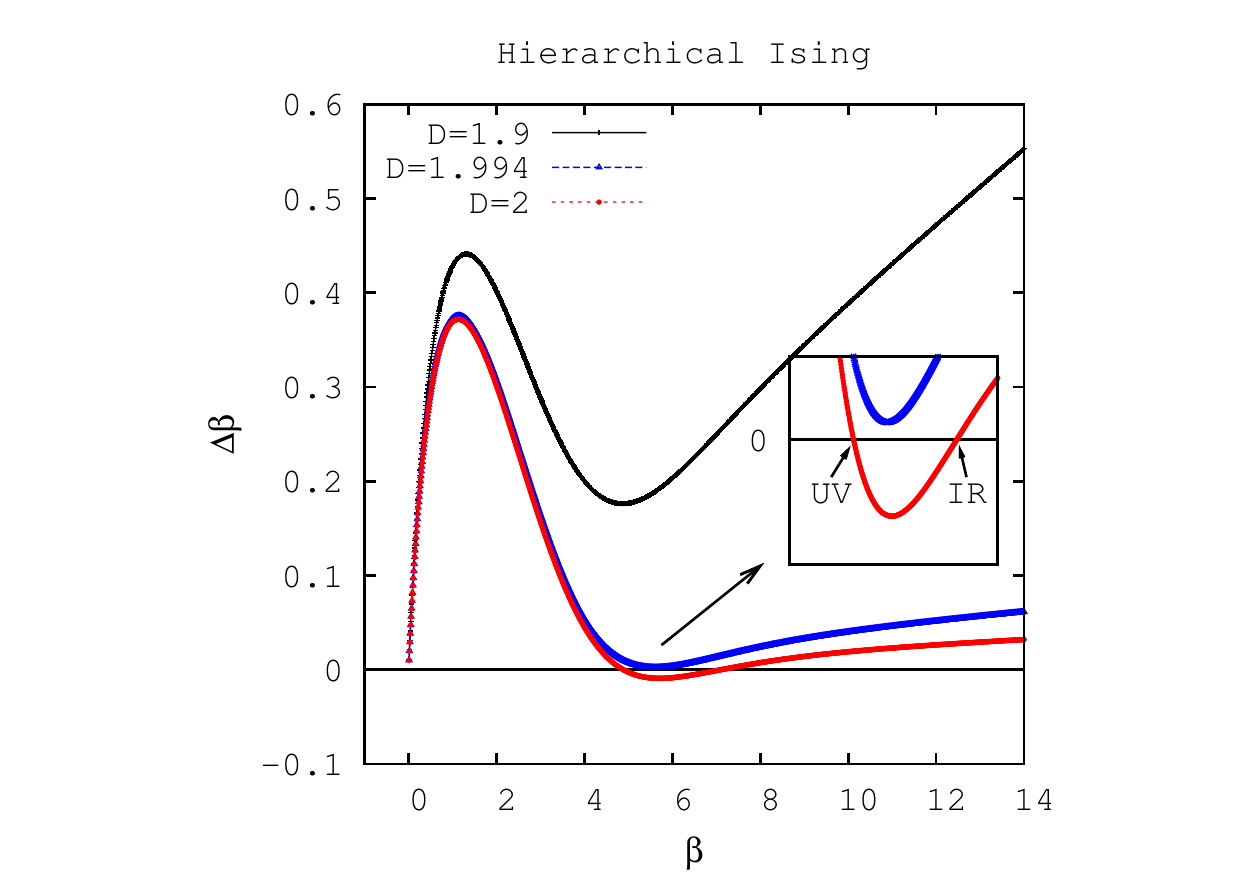}
\end{center}
\caption{Discrete $\beta$ functions for $f(n)=log(n+1)$ and $D$=1.9, 1.994, and 2 hierarchical models (dimension increases from top to bottom).}
\label{fig:dbb_hm}
\end{figure}

\begin{figure}
\begin{center}
\hspace{-0.06\textwidth}
\includegraphics[width=0.45\textwidth,keepaspectratio=]{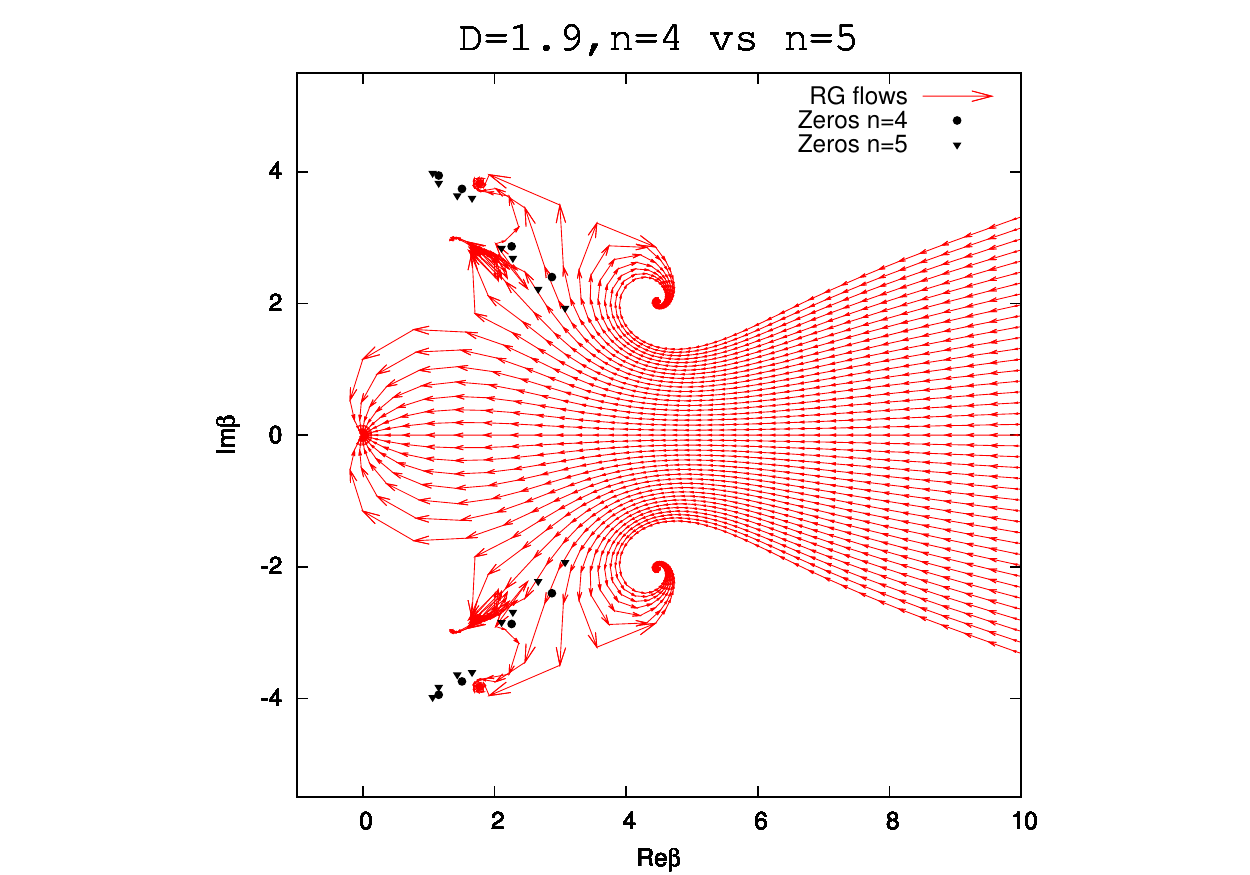}
\hspace{-0.16\textwidth}
\includegraphics[width=0.45\textwidth,keepaspectratio=]{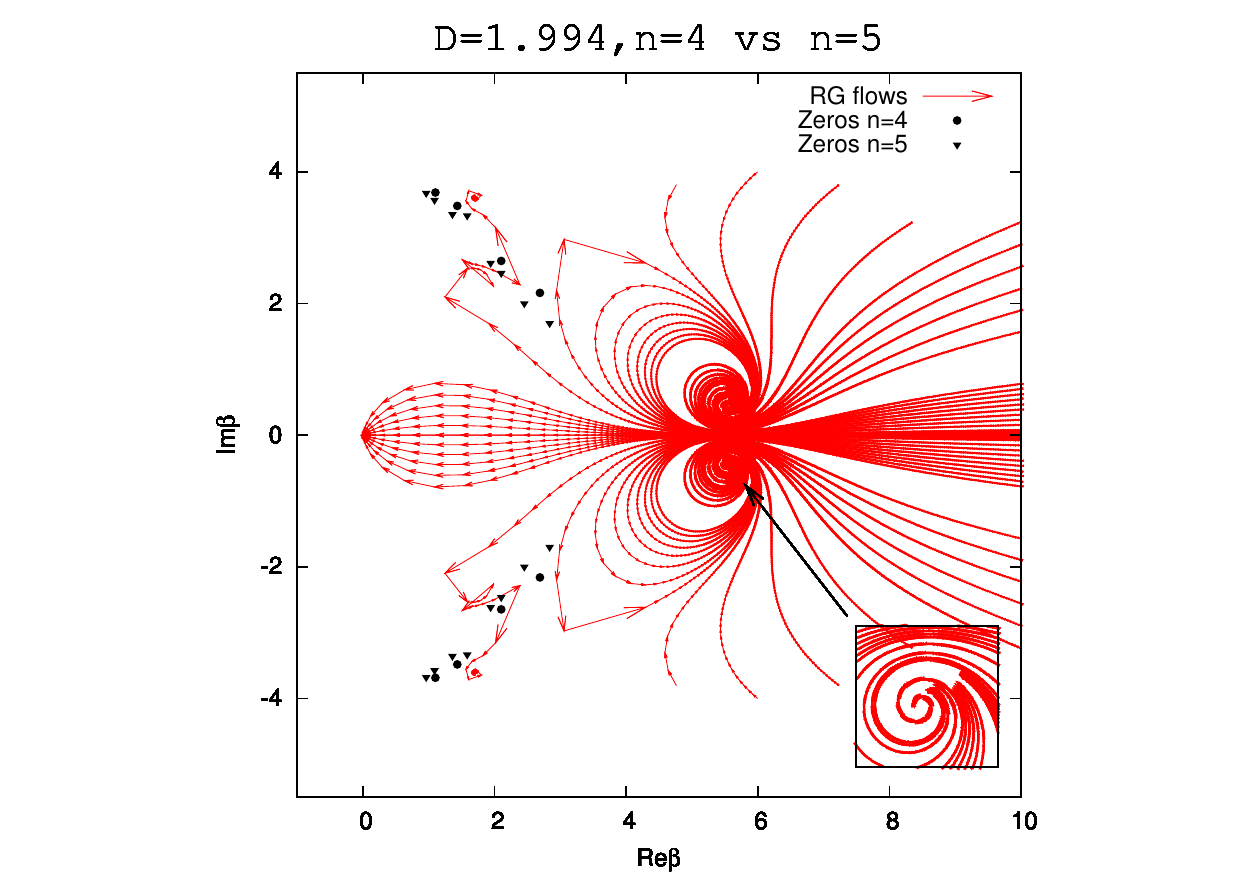}
\hspace{-0.16\textwidth}
\includegraphics[width=0.45\textwidth,keepaspectratio=]{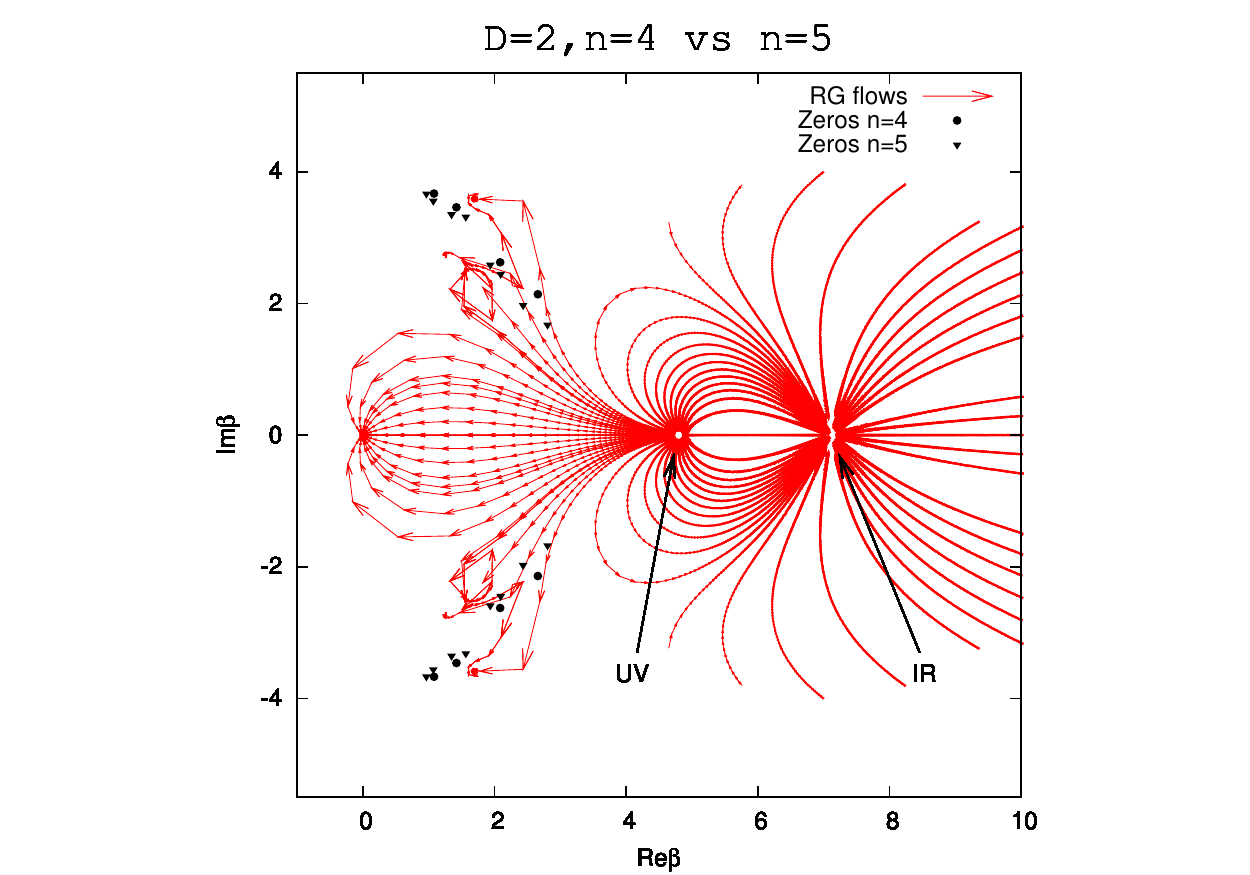}
\end{center}
\caption{Complex RG flows for $f(n)=log(n+1)$ and $D$=1.9, 1.994, and 2 hierarchical models. The flows are constructed with the two lattice matching method described in the text. }
\label{fig:flow_hm}
\end{figure}

Fixed points correspond to zeros of the $\beta$ function. It has been pointed out that zeros of the $\beta$ function can disappear in three ways as one or several parameters change \cite{PhysRevD.80.125005}. They are schematically described in Fig. \ref{fig:zeros_kaplan}. The first two ways are interchangeable depending on what parameters are used ($\alpha\leftrightarrow 1/\alpha$). For the third way, the fixed points that disappeared can be recovered in the complex parameter plane \cite{2010AnPhy.325..491M}. This third way can be easily seen for the hierarchical model by tuning the dimensionality $D$. Figure \ref{fig:dbb_hm} shows discrete $\beta$ functions for $f(n)=log(n+1)$, $D=$ 1.9, 1.994, and 2. The corresponding complex RG flows are shown in Fig. \ref{fig:flow_hm}. 

\section{Volume Effects}

\begin{figure}
\begin{center}
\hspace{-0.06\textwidth}
\includegraphics[width=0.45\textwidth,keepaspectratio=]{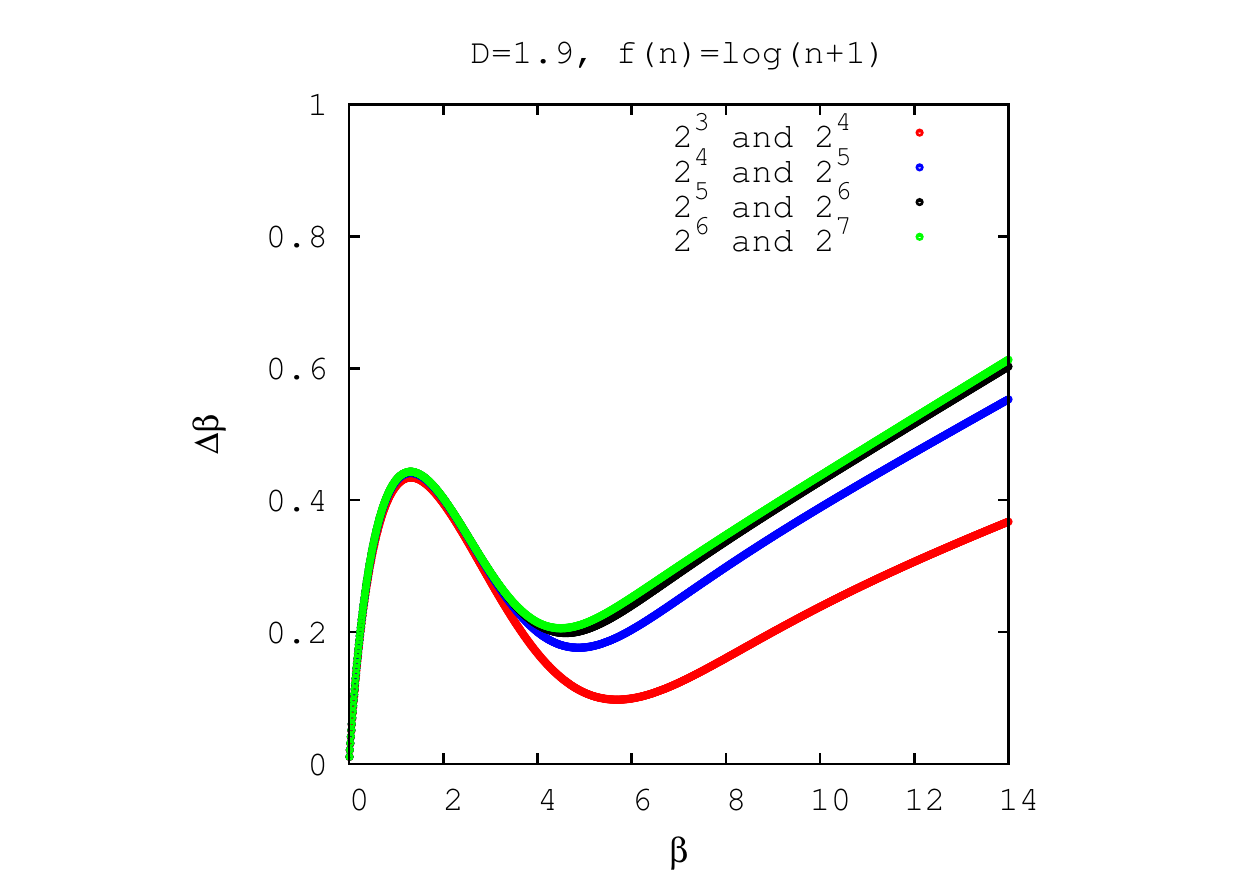}
\hspace{-0.16\textwidth}
\includegraphics[width=0.45\textwidth,keepaspectratio=]{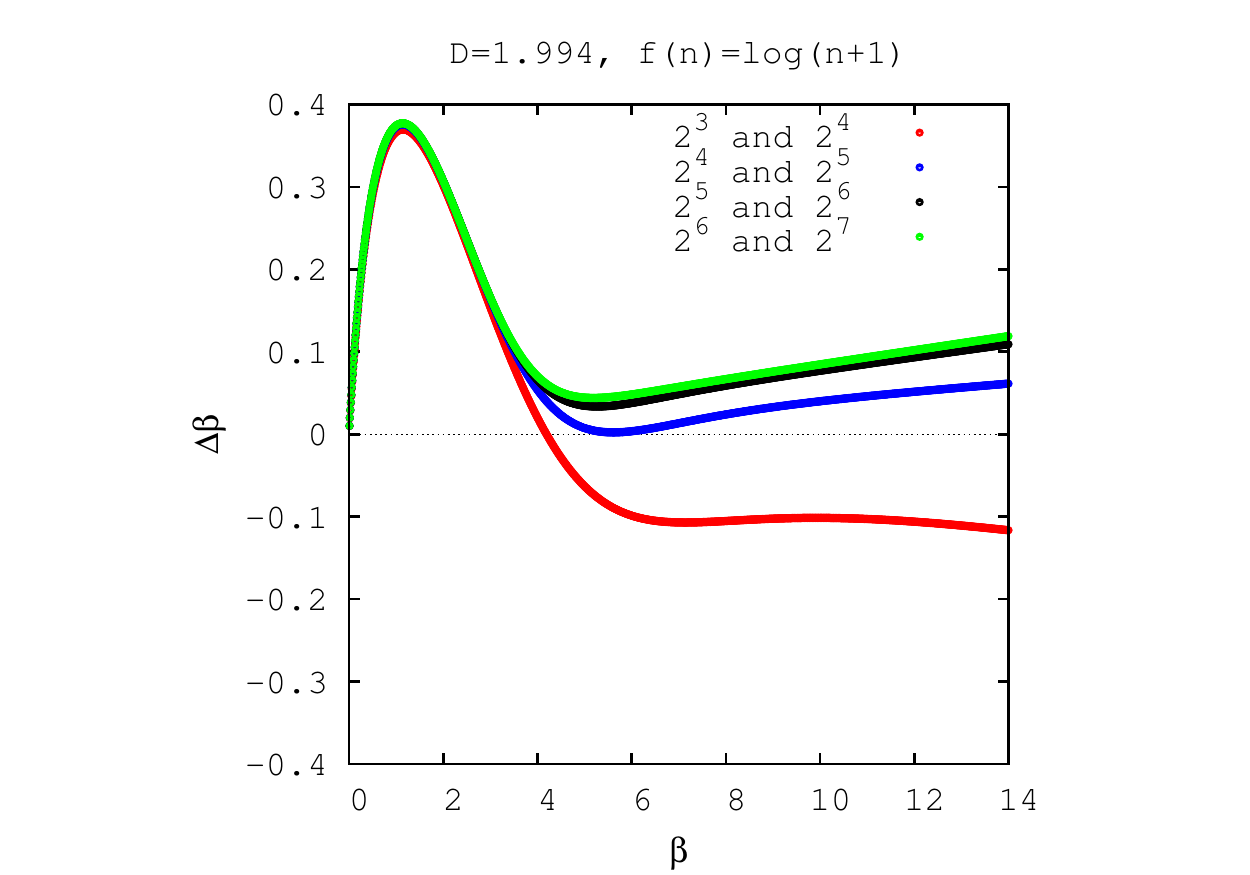}
\hspace{-0.16\textwidth}
\includegraphics[width=0.45\textwidth,keepaspectratio=]{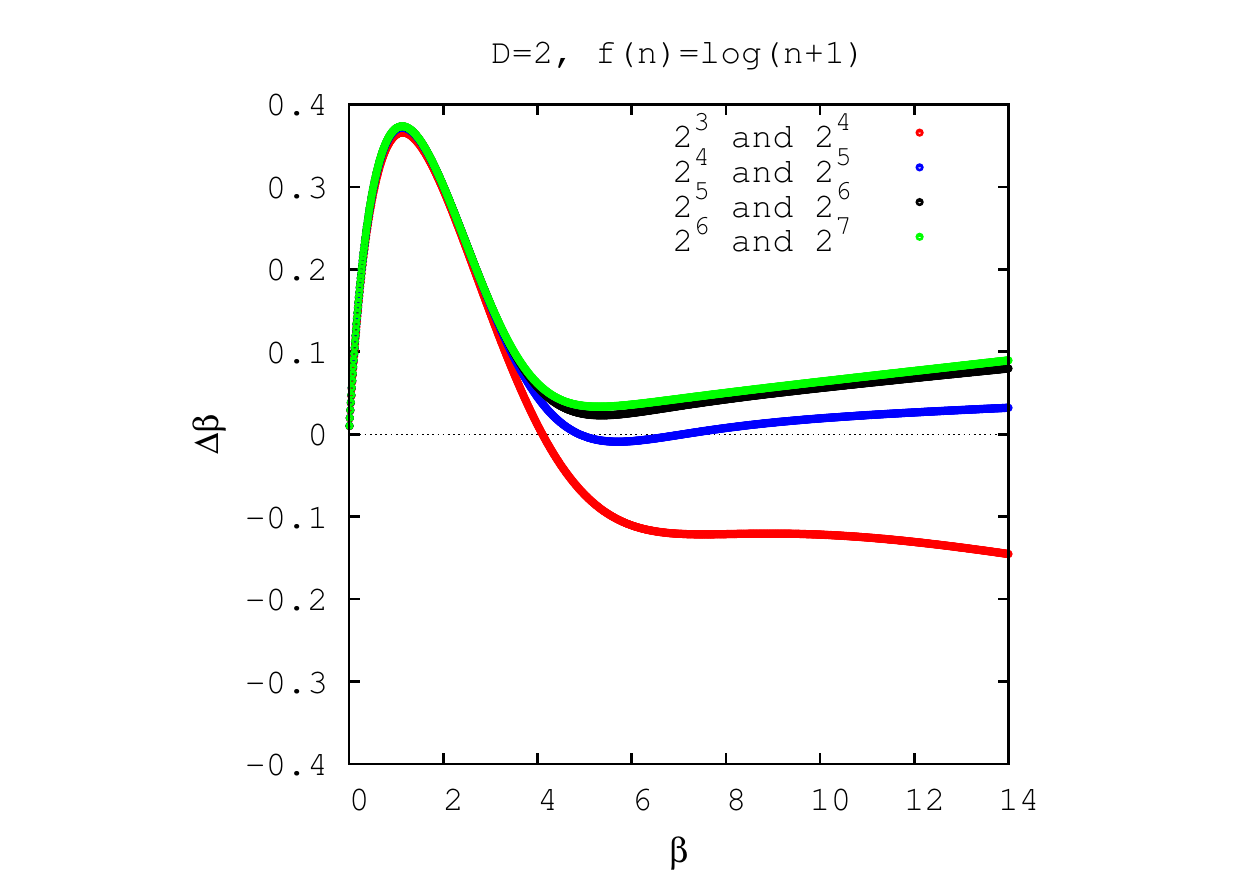}
\end{center}
\caption{Discrete $\beta$ functions for $f(n)=log(n+1)$ and $D$=1.9, 1.994, and 2 hierarchical models. Pseudo fixed points may appear for small volumes. Volume dependence is clearly shown in the figure (volume increases from bottom to top).}
\label{fig:dbb_hm_log}
\end{figure}

In \cite{PhysRevD.83.096008}, we systematically analyzed complex RG flows for the $f(n)=1$ hierarchical model and the volume dependence of Fisher's zeros (zeros of the partition function  in the complex $\beta$ plane). Fisher's zeros in lattice gauge models are discussed in \cite{PhysRevLett.104.251601}. Here we would like to illustrate volume effects in $\Delta \beta$. For this purpose, we consider the $f(n)=log(n+1)$ hierarchical model. Figure \ref{fig:dbb_hm_log} shows how discrete $\beta$ functions change with the volume. When the volume is small and the dimension parameter $D$ is large enough, pseudo fixed points may appear (Fig. \ref{fig:dbb_hm_log} middle and right). However as the volume increases, all the zeros of discrete $\beta$ functions disappear. This is expected since there is no second order phase transition for $D=2$ and $f(n)=log(n+1)$ case. Fixed points appearing in smaller volumes are finite volume artifacts.

\section{From Discrete to Continuous Flows}

Up to now, we have discussed properties of discrete $\beta$ functions. If one wants to see how the corresponding continuous Callan-Symanzik $\beta$ functions behave, we need to do some interpolations. In the following, we will construct continuous $\beta$ functions from discrete ones for $O(N)$ models. In \cite{PhysRevD.83.056009},  complex RG flows and discrete $\beta$ functions of $O(N)$ models have been constructed from both two lattice matching and a rescaling method (where the UV cutoff appearing in the dimensionless expression of the bare mass is rescaled). 
The relation between $\beta$ and $g$ is $\beta \propto g^{-2}$ for both gauge models and the $O(N)$ model. By changing the energy scale, $\beta \rightarrow \beta^\prime$ and $\beta_{CS} \rightarrow \beta_{CS}^\prime$, we get 
\begin{equation}
\beta_{CS}^\prime =\Lambda \frac{\partial}{\partial \Lambda} g^\prime =\frac{\partial g^\prime}{\partial g}( \Lambda \frac{\partial}{\partial \Lambda}g)=\frac{{g^\prime}^3}{g^3}\frac{\partial \beta^\prime}{\partial \beta}\beta_{CS}
\label{eq:bcs_bcs}
\end{equation}
A more general discussion on how to generate continuous flow from step-scaling function has been provided in \cite{PhysRevD.83.065019}. ${\partial \beta^\prime}/{\partial \beta}$ can be obtained from the discrete $\beta$ function
\begin{equation}
\frac{\partial \beta^\prime}{\partial \beta}=1-\frac{\partial \Delta \beta}{\partial \beta}
\label{eq:pbb_dbb}
\end{equation}

\begin{figure}[htp]
\begin{center}
\includegraphics[height=7cm,keepaspectratio=]{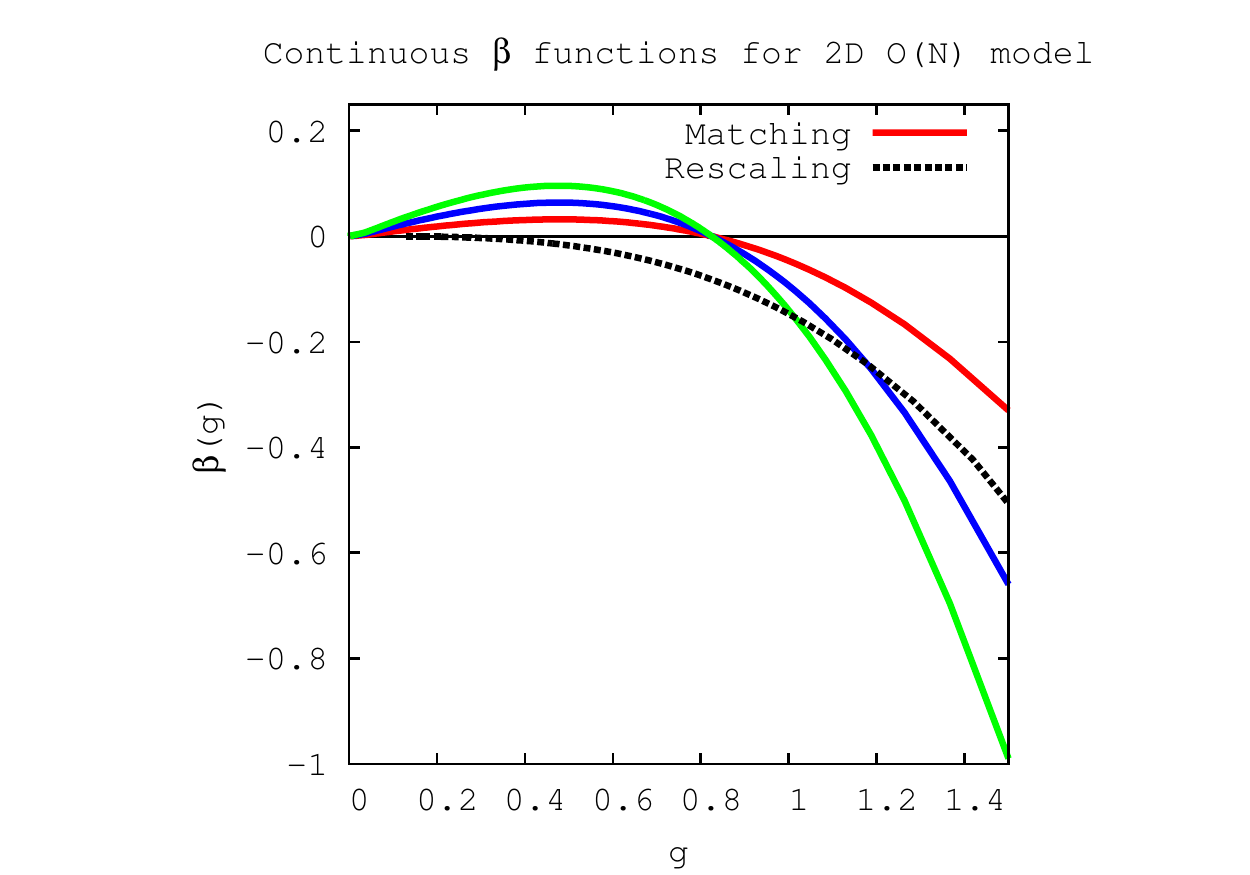}
\end{center}
\caption{Continuous $\beta$ functions constructed from both matching and rescaling discrete $\beta$ functions.}
\label{fig:cont_bcs}
\end{figure}

From Eqs. \ref{eq:bcs_bcs} and \ref{eq:pbb_dbb}, one can easily get $\beta_{CS}$ from $\Delta\beta$ by iteration. The only thing one needs to fix is the initial starting point for the interaction, i. e. ${\beta_{CS}}_0$. 
Continuous choice can be obtained by requiring that the asymptotic behavior agrees with expansions at small or large coupling. 
Figure \ref{fig:cont_bcs} shows the corresponding continuous $\beta$ functions from both matching and rescaling methods discussed in \cite{PhysRevD.83.056009}. Colored solid lines are obtained from matching discrete $\beta$ functions with different ${\beta_{CS}}_0$. A pseudo fixed point appeared in the small $g$ region and is a finite volume effect. It will go away when we further increase the volume. The black dotted line is obtained from the rescaling discrete $\beta$ function. 

\section{Conclusion}

We have successfully constructed discrete $\beta$ functions for $f(n)=log(n+1)$ hierarchical Ising models for different volumes. It has been clearly demonstrated that volume effects may generate pseudo fixed points for systems without phase transition. In order to verify whether the observed fixed point(s) is intrinsic or purely finite volume artifact, one needs to do simulations for larger volumes. An alternative way is to calculate Fisher's zeros and apply finite size scaling technique. The technique of finding Fisher's zeros from the constructed density of state has been developed for $SU(2)$ \cite{2008PhRvD..78e4503D} and tested for $U(1)$\cite{u1_misc}.  The case of $SU(3)$ with various $N_f$ is in progress. 

\begin{acknowledgments}
This research was supported in part  by the Department of Energy under Contract No. FG02-91ER40664.
\end{acknowledgments}



\begin{thebibliography}{10}

\bibitem{DeGrand:2010ba}
T.~DeGrand, ``{Lattice studies of QCD-like theories with many fermionic degrees
  of freedom},'' {\em Phil. Trans. R. Soc. A}, vol.~369, pp.~2701--2717, 2011.

\bibitem{PhysRevLett.33.244}
W.~E. Caswell, ``Asymptotic behavior of non-abelian gauge theories to two-loop
  order,'' {\em Phys. Rev. Lett.}, vol.~33, pp.~244--246, Jul 1974.

\bibitem{1974NuPhB..75..531J}
D.~R.~T. {Jones}, ``{Two-loop diagrams in Yang-Mills theory},'' {\em Nucl.
  Phys. B}, vol.~75, pp.~531--538, June 1974.

\bibitem{PhysRevD.83.096008}
Y.~Liu and Y.~Meurice, ``Lines of Fisher's zeros as separatrices for complex
  renormalization group flows,'' {\em Phys. Rev. D}, vol.~83, p.~096008, May
  2011.

\bibitem{1984PhLB..140...76H}
A.~{Hasenfratz}, P.~{Hasenfratz}, U.~{Heller}, and F.~{Karsch}, ``{Improved
  Monte Carlo renormalization group methods},'' {\em Phys. Lett. B}, vol.~140,
  pp.~76--82, May 1984.

\bibitem{1988PhLB..211..132G}
R.~{Gupta}, G.~W. {Kilcup}, A.~{Patel}, and S.~R. {Sharpe}, ``{The
  {$\beta$}-function for pure gauge SU(3)},'' {\em Phys. Lett. B}, vol.~211,
  pp.~132--138, Aug. 1988.

\bibitem{2007JPhA...40...39M}
Y.~{Meurice}, ``{Nonlinear aspects of the renormalization group flows of
  Dyson's hierarchical model},'' {\em J. Phys. A}, vol.~40, p.~39, June 2007.

\bibitem{1969CMaPh..12...91D}
F.~J. {Dyson}, ``{Existence of a phase-transition in a one-dimensional Ising
  ferromagnet},'' {\em Comm. Math. Phys.}, vol.~12, pp.~91--107, June 1969.

\bibitem{1971CMaPh..21..269D}
F.~J. {Dyson}, ``{An Ising ferromagnet with discontinuous long-range order},''
  {\em Comm. Math. Phys.}, vol.~21, pp.~269--283, Dec. 1971.

\bibitem{pool1972mathematical}
F.~J. {Dyson}, { Existence and nature of phase transitions in
  one-dimensional Ising ferromagnets}.
\newblock SIAM-AMS proceedings, American Mathematical Society, 1972.

\bibitem{PhysRevD.80.125005}
D.~B. Kaplan, J.-W. Lee, D.~T. Son, and M.~A. Stephanov, ``{Conformality
  lost},'' {\em Phys. Rev. D}, vol.~80, p.~125005, Dec 2009.

\bibitem{2010AnPhy.325..491M}
S.~{Moroz} and R.~{Schmidt}, ``{Nonrelativistic inverse square potential, scale
  anomaly, and complex extension},'' {\em Ann. Phys.}, vol.~325, pp.~491--513,
  Feb. 2010.

\bibitem{PhysRevLett.104.251601}
A.~Denbleyker, D.~Du, Y.~Liu, Y.~Meurice, and H.~Zou, ``Fisher's zeros as the
  boundary of renormalization group flows in complex coupling spaces,'' {\em
  Phys. Rev. Lett.}, vol.~104, p.~251601, Jun 2010.

\bibitem{PhysRevD.83.056009}
Y.~Meurice and H.~Zou, ``{Complex renormalization group flows for $2D$
  nonlinear $O(N)$ sigma models},'' {\em Phys. Rev. D}, vol.~83, p.~056009, Mar
  2011.

\bibitem{PhysRevD.83.065019}
T.~L. Curtright and C.~K. Zachos, ``Renormalization group functional
  equations,'' {\em Phys. Rev. D}, vol.~83, p.~065019, Mar 2011.

\bibitem{2008PhRvD..78e4503D}
A.~{Denbleyker}, D.~{Du}, Y.~{Liu}, Y.~{Meurice}, and A.~{Velytsky}, ``{Series
  expansions of the density of states in $SU(2)$ lattice gauge theory},'' {\em
  Phys. Rev. D}, vol.~78, p.~054503, Sept. 2008.

\bibitem{u1_misc}
A.~Bazavov, B.~A. Berg, D.~Daping, and Y.~Meurice, ``{Density of states and
  Fisher's zeros in compact $U(1)$ pure gauge theory},'' preprint in progress.

\end{thebibliography}
\end{document}